\documentclass{article}

\usepackage[top=2cm, bottom=2cm, left=2cm, right=2cm]{geometry}
\usepackage{graphicx}
\usepackage{multirow}
\usepackage{amsmath,amssymb,amsfonts}
\usepackage{amsthm}
\usepackage{mathrsfs}
\usepackage{xcolor}
\usepackage{textcomp}
\usepackage{manyfoot}
\usepackage{booktabs}
\usepackage{algorithm}
\usepackage{algorithmicx}
\usepackage{algpseudocode}
\usepackage{listings}
\usepackage{caption}

\begin {document}
\date{}
\title{Multiagent Copilot Approach for Shared Autonomy between Human EEG and TD3 Deep Reinforcement Learning}

\author{
Chun-Ren Phang\footnote{Department of Electrical and Mechanical Engineering, Nagoya Institute of Technology, Nagoya, Aichi, Japan.} \footnote{Center of Biomedical Physics and Information Technology, Nagoya Institute of Technology, Nagoya, Aichi, Japan. \\\indent\hspace{0.1cm}(e-mail: phangcr@gmail.com; ahirata@nitech.ac.jp)}\hspace{0.2cm}, 
Akimasa Hirata \textsuperscript{*$\dagger$}\, 
} 

\maketitle

\abstract{Deep reinforcement learning (RL) algorithms enable the development of fully autonomous agents that can interact with the environment. Brain--computer interface (BCI) systems decipher human implicit brain signals regardless of the explicit environment. In this study, we integrated deep RL and BCI to improve  beneficial human interventions in autonomous systems and the performance in decoding brain activities by considering environmental factors. Shared autonomy was allowed between the action command decoded from the electroencephalography (EEG) of the human agent and the action generated from the twin delayed DDPG (TD3) agent for a given environment. Our proposed copilot control scheme with a full blocker (Co-FB) significantly outperformed the individual EEG (EEG-NB) or TD3 control. The Co-FB model achieved a higher target approaching score, lower failure rate, and lower human workload than the EEG-NB model. The Co-FB control scheme had a higher invisible target score and level of allowed human intervention than the TD3  model. We also proposed a disparity $d$-index to evaluate the effect of contradicting agent decisions on the control accuracy and authority of the copilot model. We found a significant correlation between the control authority of the TD3 agent and the performance improvement of human EEG classification with respect to the $d$-index. We also observed that shifting control authority to the TD3 agent improved performance when BCI decoding was not optimal. These findings indicate that the copilot system can effectively handle complex environments and that BCI performance can be improved by considering environmental factors. Future work should employ continuous action space and different multi-agent approaches to evaluate copilot performance. 
}

\vspace{0.1in}

\section{Introduction}\label{sec1}

\subsection{Brain--Computer Interfaces}\label{sec1.1}

Neuroimaging advancement has enabled vivid visualization of the human brain in both structural and functional contexts. The brain--computer interface (BCI) was developed using neuroimaging modalities such as electroencephalography (EEG) to enable direct brain--computer communication. The decoding of brain patterns can be used to control cursors \cite{Li2010}, \cite{Long2011}, \cite{Allison2012}, wheelchairs \cite{Li2013}, \cite{Carlson2013} or exoskeletons \cite{Frisoli2012}, \cite{Barsotti2015}, \cite{Frolov2016}. 

P300 oddball stimulus, steady-state visually evoked potential (SSVEP), and motor-related somatosensory rhythms are the three neurological signals extensively used for BCI commands. Both SSVEP- and P300-based BCIs require a short training time and have shown promising outcomes. However, both systems require exogenous stimulation. Conversely, motor-related brain rhythms, generated from motor imagery (MI) or motor planning, involve mentally simulating a motor task without physically engaging the muscles \cite{Decety1990}. According to \cite{Nielsen2006}, MI can generate movement-related cortical potential similar to that of actual muscle action without the requirement of external stimulus. EEG features such as event-related desynchronization \cite{Liu2019}, beta rebound \cite{Sun2013, Hashimoto2013}, empirical mode decomposition \cite{Mohamed2018}, filter bank common spatial pattern (FBCSP) \cite{ang2008}, and functional connectivity (FC) \cite{phang2022} have been adopted to detect motor-related brain activities. 

Most users can now engage with BCIs with promising performance because of advancements in BCI research. However, despite the tremendous effort of training, a non-negligible portion of the population has failed to operate one or more types of BCI systems \cite{Blankertz2010}. According to estimates by \cite{Vidaurre2010}, approximately 15 \%--30 \% of the population is unable to generate brain signals that can be translated into BCI commands. Nevertheless, conventional BCIs decode implicit brain signals without considering the interacting environment. 

\subsection{Human-In-The-Loop}\label{sec1.2}

Recent research has focused on implementing a human-in-the-loop system that allows feedback from humans to facilitate the training of deep reinforcement learning (RL). Human intervention is important to fine-tune the deep RL agent to behave more naturally and in the presence of safety constraints \cite{bharadhwaj2020}. A human--artificial intelligence copilot optimization (HACO) algorithm was proposed to allow human experts to take control in a risky environment and demonstrate safe control \cite{li2022}. The deep RL agent trained with HACO achieved safer driving than conventional RL baselines. In addition, the proposed DQN-TAMER denotes human facial expressions, decoded with a convolutional neural network, as the human reward for deep Q-learning (DQN). 

Moreover, the BCI system developed with EEG brain signals was adopted as implicit human feedback during the learning phase of DQN \cite{xu2021}. The error-related potential (ErrP) can be measured from the scalp when humans are presented with erroneous stimuli, such as incorrect actions taken by the DQN agent. The ErrP is useful for providing direct brain-to-RL interactions to prevent potentially dangerous actions. Xu et al. demonstrated that the training step of the proposed ErrP-DQN model can be significantly reduced compared with that of the conventional DQN model \cite{xu2021}. However, although these studies allowed a certain extent of human involvement during deep RL training, how humans could copilot with deep RL during the task execution (testing) phase was not demonstrated. 

\subsection{Motivations and Objectives}\label{sec1.3}

A recent survey discussed the difficulty in controlling machines by nonexpert users and urged increased human-centered research to improve the interface of machine control \cite{rea2022}. The three main strategies for cooperation between humans and machines include human-dominant, machine-dominant, and human--robot consensus \cite{yang2021, flemisch2019}. Haptic shared controls have been explored in the context of automated vehicles \cite{schwarting2017, mars2014} and drones \cite{reddy2018}. Recently, ErrP--BCI was incorporated with an actor--critic (A2C) deep RL algorithm to navigate in environments with or without obstacles \cite{wang2022}. This allowed shared autonomy between the human brain and deep RL agent not in the training phase but during the task execution phase. The results showed that the ErrP--A2C shared autonomy model could complete navigation tasks with high efficiency. The authors suggested that the shared autonomy control scheme could effectively deal with the uncertainty in human EEG feedback. Similarly to \cite{xu2021}, because ErrP is elicited in the presence of an erroneous stimulus, human users can only passively provide “pass” and “stop” input commands. 

Therefore, in this study, we present an active BCI--deep RL shared autonomy control scheme based on a multi-agent copilot approach to improve the performance of both human and deep RL agents. We hypothesized that active copilots could significantly improve task performance in complex environments while simultaneously reducing human workload.

\section{Methods}\label{sec2}

\subsection{Electroencephalography Datasets}\label{sec2.1}

In this study, we mainly used the EEG dataset presented by \textit{Kaya et al.} \cite{kaya2018}, which comprised EEG recordings from 12 healthy participants. The EEG signals were acquired using the EEG-1200 JE-921A system, following the standard 10/20 international configuration. The sampling frequency was set at 200 Hz, and the impedance remained consistently below 10 k$\Omega$ during the experimental sessions. Grounding was achieved using two earbud electrodes, A1 and A2, while “System 0 V” served as the virtual reference point, defined as 0.55 × (C3 + C4)V in the EEG-1200 technical manual. The EEG channels included Fp1, F3, F7, C3, P3, T3, T5, Fz, Cz, Pz, Fp2, F4, F8, C4, P4, T4, T6, O1, and O2. To mitigate electrical noise, a 50-Hz notch filter was applied to the recorded signal. In this study, we selected the six-class HaLT dataset, which encompassed the left-hand, right-hand, left-foot, right-foot, tongue, and rest tasks. During recording, a stimulus was displayed on the screen for 1 s, prompting subjects to mentally imagine the corresponding motor movement. Following each trial, subjects were instructed to focus on a fixation point for 1.5--2.5 s. Each subject completed 143--156 trials for each task. To control the four-directional player agent, the left-hand, right-hand, left-foot, and right-foot tasks were chosen as control commands from all subjects.

The proposed model was also assessed using the BCI Competition dataset \cite{tangermann2012}. The EEG data were acquired using 22 Ag/AgCl EEG channels (C5, C3, C1, Cz, C2, C4, C6, Fz, FC3, FC1, FCz, FC2, FC4, CP3, CP1, CPz, CP2, CP4, P3, Pz, P4, and Oz) from nine healthy participants. Signals were recorded with reference to the left mastoid, and the right mastoid served as the ground. A sampling rate of 250 Hz was used, and the data were bandpass filtered between 0.5 and 100 Hz. A 50-Hz notch filter was used to eliminate line noise. This dataset consisted of 288 trials, with 72 recordings per task. During the recording sessions, a stimulus appeared on the screen, prompting the participants to mentally imagine left-hand, right-hand, left-foot, or right-foot motor movements within a 3-s interval. Following each trial, the participants were instructed to maintain their gaze on a fixation point for 2 s. All four imagery tasks were selected as control commands for the four-directional player agent.

\subsection{Electroencephalography Analysis}\label{sec2.2}

Weighted FC was calculated by determining the pairwise Pearson’s correlation between EEG signal pairs. The correlation coefficient, denoted as $r_{xy}$, quantifies the degree of connectivity between the EEG signals acquired from channels $x$ and $y$ (as shown in Eq. \ref{pearcor}). The combination of these connections across all signal pairs resulted in a symmetric connectivity matrix of dimensions $N_c \times N_c$, where $N_c$ represents the total number of channels.
\begin{equation} \label{pearcor}
r_{xy} = \dfrac{\sum_{i=1}^{N_t} (x_i - \bar{x})(y_i - \bar{y})}{\sqrt{\sum_{i=1}^{N_t} (x_i - \bar{x})^2}\sqrt{\sum_{i=1}^{N_t} (y_i - \bar{y})^2}}
\end{equation} where $N_t$ represents the number of time points, and $\bar{x}$ and $\bar{y}$ denote the mean EEG potentials from channels $x$ and $y$, respectively.

BP is a widely used measure to understand EEG activities. We computed the BPs among delta (1--4 Hz), theta (4--7 Hz), alpha (8--13 Hz), beta (14--30 Hz), and gamma (30--100 Hz) frequency ranges for the detection of four-class MI tasks. The power of the signal within specific frequency ranges can be computed using a periodogram (Eq. \ref{bandpower}). For a signal $x_{n}$ sampled at a $\Delta t$ sampling interval, the power can be defined as follows:
\begin{equation} \label{bandpower}
\hat{P} (f)=\Bigg|\sum_{n=0}^{N-1}x_{n}e^{-i2\pi f \Delta tn}\Bigg|^2
\end{equation}

The extracted $N \times N$ FC and $5 \times N$ BP features were fed into a multiclass LDA classifier to classify the four motor imagination tasks. LDA was adopted in our proposed model because it is a well-established and computationally efficient classifier. The multiclass LDAs were evaluated using 10-fold cross-validation, with 90\% training and 10\% testing trials. The posterior probability and predicted label of the testing trials were aggregated across each fold as the control command of the player agent. The random guessing threshold for the four-class classification was 25\%. 

\subsection{Twin Delayed DDPG Deep Reinforcement Learning}\label{sec2.3}

\begin{figure*}[h]%
\centering
\captionsetup{width=0.9\textwidth}
\includegraphics[width=0.8\textwidth]{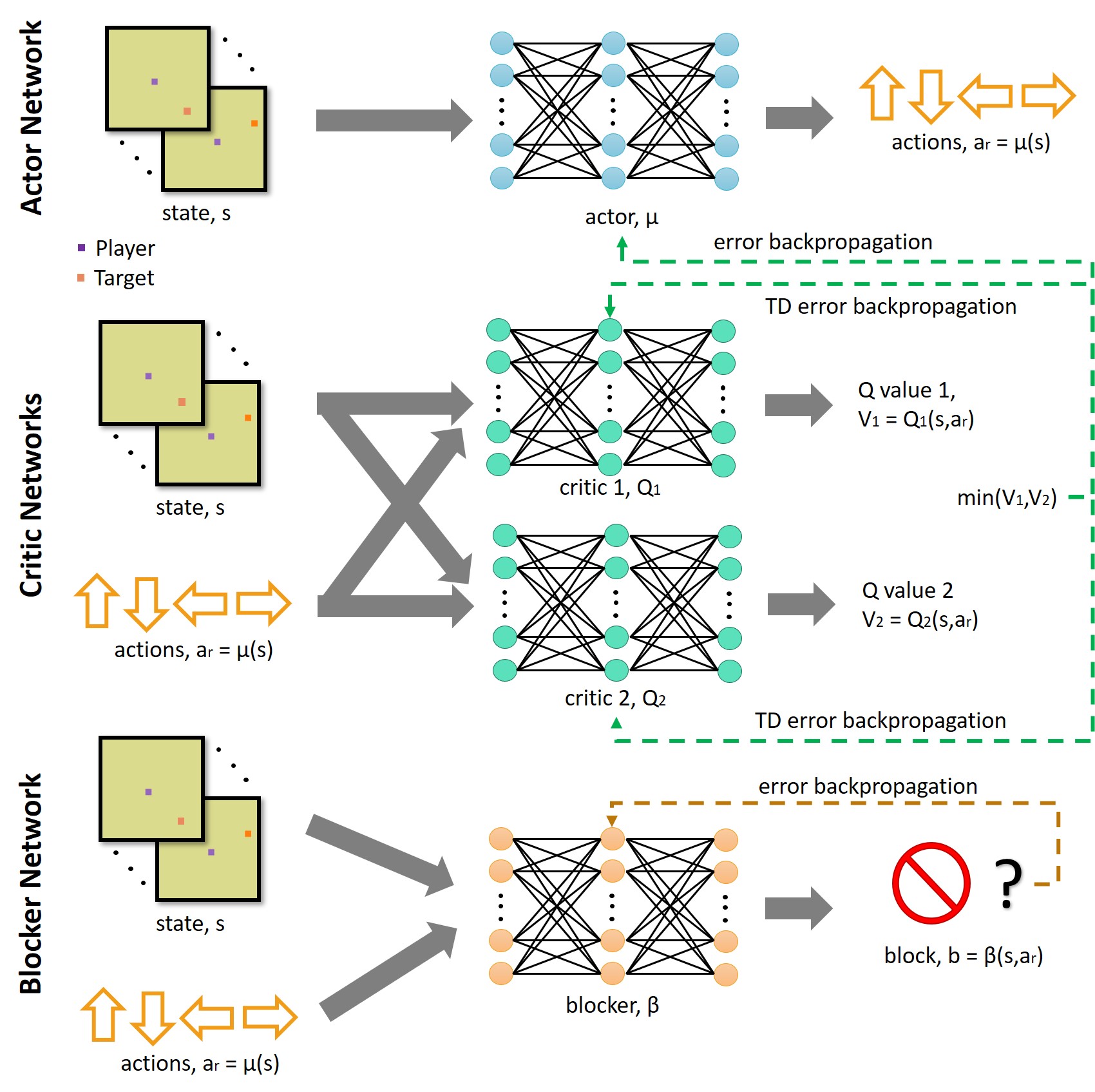}
\caption{Schematic illustrating the training of the network and blocker networks. The TD3 algorithm comprises two critic networks, one actor network, and their duplicated target networks. For simplification, the target networks were not visualized. A blocker network was also incorporated to safeguard risky actions due to misclassification of human EEG.}\label{RL_training}
\end{figure*}

TD3 deep RL with a blocker was implemented as one of the major subsets of the proposed copilot model. TD3 was originally proposed in \cite{fujimoto2018} to deal with the overestimation bias in the A2C network. The TD3 algorithm comprises two critic networks, one actor network, and their duplicated target networks. A blocker network \cite{saunders2017, prakash2019} was also embedded to safeguard against risky actions caused by the misclassification of human EEG intentions. The TD3 algorithm was trained using experience replay. A simplified training flowchart of the TD3 and blocker networks is shown in Figure \ref{RL_training}. The value function of the critic networks can be trained using temporal difference learning with an error function as follows \cite{van2016}: 
\begin{equation}
\delta_{c} = \frac{1}{M}\sum^{M}_{t=1}(r_{t} + \gamma Q(s_{t+1},a_{t+1}) – Q(s_{t},a_{t}))^2
\end{equation}
\noindent where $M$ represents the size of the minibatch, $r_{t}$ is the reward from the environment after performing an action, and $r_{t} + \gamma Q(s_{t+1},a_{t+1})$ denotes the temporal target. The TD3 algorithm introduced a clipped double Q-learning method to reduce the overestimation error, where the minimum $Q(s_{t+1},a_{t+1})$ between two critic networks was selected as the target $Q-$value. The actor network is trained using a deterministic policy gradient algorithm, where the network weights are trained to maximize the expected return $Q(s_{t},a_{t})$ of taking action $a_{t}$ in state $s_{t}$. The loss of the actor network was backpropagated from the critic network. 
\begin{equation}
\delta_{a} = \frac{1}{M}\sum^{M}_{t=1}(-Q(s_{t},\mu(s_{t})))
\end{equation}
\noindent To maintain the stability of the target networks, a delayed policy update approach was adopted to update the target actor and critic networks, with an update weight of $\tau$
\begin{equation}
\phi’ = \tau\phi + (1 - \tau)\phi’
\end{equation}
\noindent where $\phi’$ denotes the weight of the target network and $\phi$ denotes the trained weight of the non-target network. The blocker network was trained with a root mean squared error between the expected risk $b = \beta(s_{t},a_{t})$ and the $r_{g}$ failed reward from the environment.
\begin{equation}
\delta_{b} = \frac{1}{M}\sum^{M}_{t=1}(\beta(s_{t},a_{t}) – r_{g,t})^2
\end{equation}

Both actors, critics, and blockers comprised 1 input layer, 2 hidden layers with 16 and eight nodes, respectively, and 1 output layer. The input for the actor was the minimally crafted states, whereas the input for the critic and blocker was both the states and actions taken by the actor. All three networks were optimized using Adam with a learning rate of 0.0003 \cite{kingma2014}. 

\begin{figure*}[h]%
\centering
\captionsetup{width=0.9\textwidth}
\includegraphics[width=0.7\textwidth]{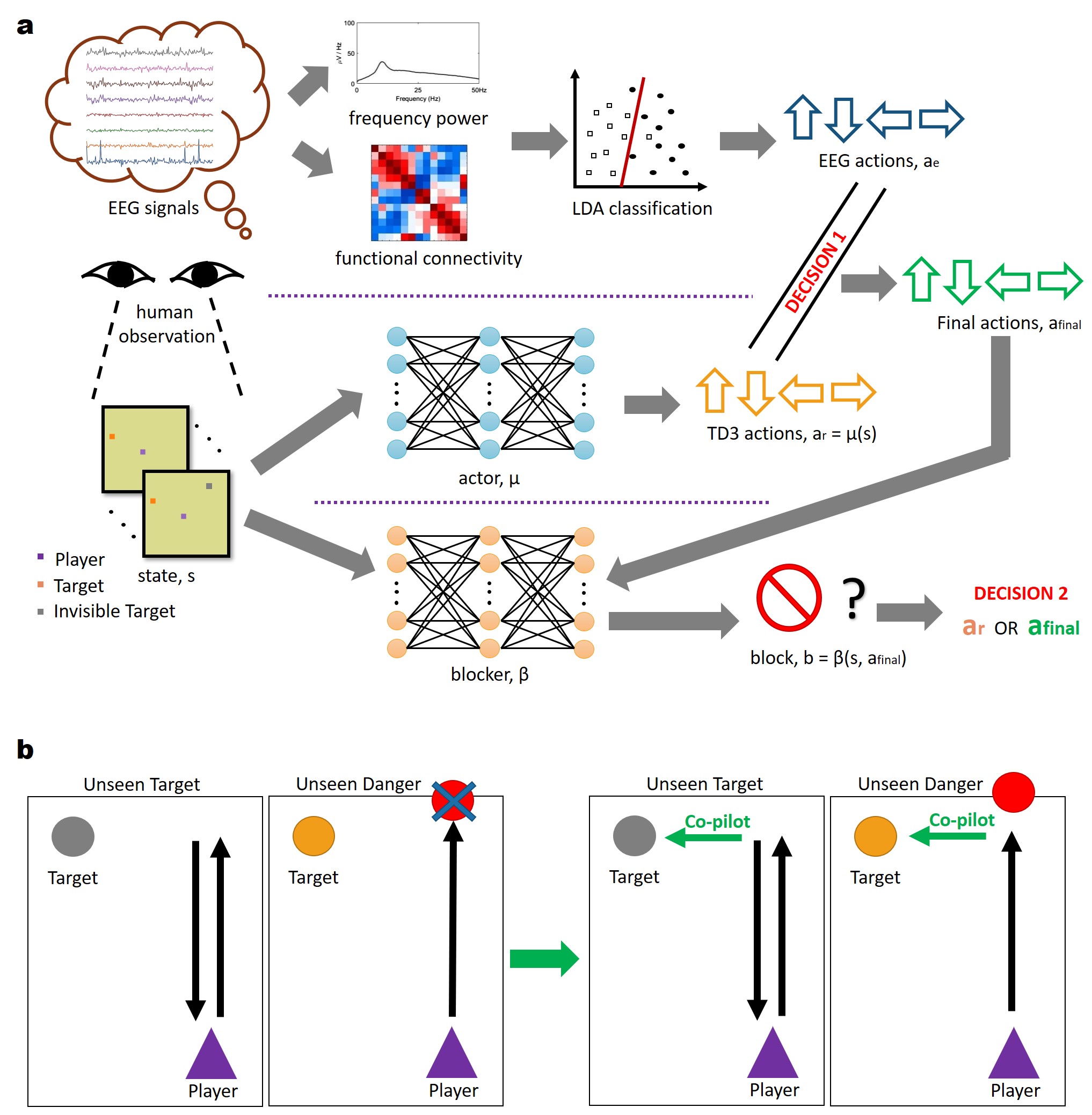}
\caption{(a) Multi-agent copilot approach for generating control actions. The human agent (EEG activities), trained TD3 agent, and trained blocker agent performed their respective actions after evaluating the environment. The copilot model aggregates inputs from all three agents, and a final action command is generated to control the player agent. (b) Advantages of the proposed multi-agent copilot control scheme in complex environments. The copilot scheme could reduce the risk of danger that is invisible to one agent and improve the target approaching ability when the target is invisible to one agent. }\label{copilot}
\end{figure*}

\subsection{Copilot Scheme}\label{sec2.4}

Figure \ref{copilot} (a) illustrates the general concept of the TD3-EEG copilot. A final decision could be made after considering input from the human, TD3, and blocker agents. A copilot control scheme was implemented using the decision tree approach. In the first layer, the human actions decoded from the EEG features (human agent), including FC ($a_{e,fc}$) and BP ($a_{e,bp}$), were compared. In the case where both classifications agree with each other $a_{e,fc} = a_{e,bp}$, a temporary final action can be computed as $a_{final} \Leftarrow a_{e,fc}$ or $a_{final} \Leftarrow a_{e,bp}$. Conversely, in the case where the EEG classifications disagree with each other $a_{e,fc} \neq a_{e,bp}$, both $a_{e,fc}$ and $a_{e,bp}$ would be compared with the action generated from the TD3 agent ($a_{r}$). The temporary final action $a_{final}$ could be appointed as $a_{final,1} \Leftarrow a_{r}$, if $a_{r} \Leftarrow a_{e,fc}$ or $a_{r} \Leftarrow a_{e,bp}$. However, when neither of the above-mentioned criteria was met ($a_{e,fc} \neq a_{e,bp}$, $a_{r} \neq a_{e,fc}$, and $a_{r} \neq a_{e,bp}$), the LDA posterior probabilities of $a_{e,fc}$ and $a_{e,bp}$ would be fused by averaging, and the temporary final action $a_{final}$ was assigned as the action decoded from the fused posterior probabilities ($a_{final} \Leftarrow a_{e,pp}$). In the last layer, a final decision is made by the blocker ($\beta$) to determine the risk $b$ of performing the temporary final action $a_{final}$. The temporary final action $a_{final}$ would be performed if no risk is detected by the blocker ($b = 0$). In a situation where risky actions were detected ($b = 1$), the temporary final action $a_{final}$ would be stopped, and the TD3 actions $a_{r}$ would be performed instead, thereby reducing the risk caused by the misclassification of human EEG features. The advantages of the proposed multi-agent copilot control schemes in complex environments are shown in Figure \ref{copilot} (b). 

\subsection{Simulation Environment}\label{sec2.5}

Two target approaching environments were simulated to evaluate the performance of player agents using different control schemes. Both simulated $16 \times 16$ environments consisted of a player agent and randomly appearing targets. The targets appeared once, and the next target was prompted after the player agent approached the prior target. One of the environments contained additional targets that were invisible to the TD3 agent but visible to the human agent. The invisible targets had an appearance probability of 0.01, and similar to the visible targets, the subsequent invisible target was only prompted after the prior invisible target was approached. The invisible targets could represent unseen states during TD3 training or undetectable states (due to reasons such as a narrow observation field or suboptimal sensor location). The targets were invisible to the TD3 agent but could easily be noticed by the human agent. In this case, human intervention is required to navigate the environment. A demonstration of the simulated environment is shown in Figure ~\ref{scores}. In every new episode, the player agent spawns in the center of the environment, and targets are randomly spawned in other regions. The player agent was allowed to perform four commands, left, right, up, and down, to approach the targets. Every command moves the player agent one step forward. The player agent would be awarded a score of 1 when it approached a visible target, and a score of 10 would be awarded for each approached invisible target. To facilitate convergence, we devised non-target rewards during TD3 training. A reward of $r=0.5$ was given when the player agent moved closer to the target, and $r=-0.5$ was provided when the agent moved further away from the target. A penalty score of  10 is given if the target player moves across the boundary of the environment (fail in the game), and the episode ends. We minimally crafted the states to reduce the input dimension fed to train the TD3 agent. Four-dimensional states, including the horizontal and vertical locations of the player agent and target, were generated as the input environment. 

\subsection{Computation of Accuracy and Authority}\label{sec2.6}

To further evaluate the performance of the proposed Co-FB model, the control authority and classification accuracy were computed. Here, we also introduce a new term, $disparity, d$, which represents the rate of disparity $d \in [0, 1]$ between TD3 actions $a_{r}$ and human actions $a_{e}$. In this study, $d = 0$ denotes full agreement between the two action sets $A_{e} = A_{r}$, whereas $d = 1$ indicates that both action sets do not agree with each other $A_{e} \cap A_{r} = \emptyset$. Considering $disparity, d$, we can evaluate the proposed model on the basis of different situations. The $disparity$ index could be affected by several factors such as the environment, emergency, accuracy of EEG classification, and individual habits. In this study, we did not investigate how the $d$-index could be varied by different factors. Instead, we allow the $d$-index to vary from 0 to 1 and discuss how this can affect the control authority and classification accuracy of the Co-FB model. 

The control authorities of human EEG ${Ath}_{e}$ and TD3 RL ${Ath}_{r}$ can be computed from the copilot scheme algorithm as follows: 
\begin{equation}
{Ath}_{e} = 1 - W_4,r
\end{equation}
\begin{equation}
{Ath}_{r} = W_{1,r} + W_{2,r} + W_{3,r} + W_{4,r}
\end{equation}
\noindent where $W$ represents the probability of actions, weighted with the $disparity$ index $d$. $W_{r}$ and $W_{e}$ represent the weighted probability of TD3 action $a_{r}$ and human action $a_{e}$, respectively. The $W$ of the first decision layer, can be calculated as follows:
\begin{equation}
W_{1,e} = P(a_{e,fc} \cap a_{e,bp})
\end{equation}
\begin{equation}
W_{1,r} = W_{1,e} \times (1 - d)
\end{equation}
\noindent $W_{1,r}$ was computed as the probability of $a_{r} = a_{e,fc} = a_{e,bp}$. The subsequent $W_{2,e}$ and $W_{2,r}$ in decision layer 2, can be calculated by multiplying the rate of agreement $(1-d)$ by the probability of $a_{e,fc} \neq a_{e,bp}$:
\begin{equation}
W_{2} = (1 - d) \times ({1 - W}_{1,e})
\end{equation}
\begin{equation}
W_{2} = W_{2,e} = W_{2,r}
\end{equation}
\noindent The authorities of the TD3 and human agents were the same as both agents shared the same actions. From decision layer 3, $W_{3,e}$ was computed by multiplying the disparity index $d$ with the probability of $a_{e,fc} \neq a_{e,bp}$. $W_{3,r}$ denotes the rate of agreement $(1 - d)$ between $a_{r}$ and $a_{e,pp}$ 
\begin{equation}
W_{3,e} = d \times ({1 - W}_{1,e})
\end{equation}
\begin{equation}
W_{3,r} = W_{3,e} \times (1 - d)
\end{equation}
\noindent The authorities in the final layer are indicated as the product of the probability of human actions and the probability of risky actions, as follows:
\begin{equation}
W_{4,r} = P(b = 1) \times (W_{1,e} + W_{3,e})
\end{equation}
\begin{equation}
W_{5,e} = P(b \neq 1) \times (W_{1,e} + W_{3,e})
\end{equation}
\noindent where $b$ represents the blocker decision $\beta(s,a_{final})$.

With the computed control authority, the classification accuracy of human EEG in Co-FB, ${Acc}_{c}$, was calculated as follows:
\begin{equation}
\begin{aligned}
{Acc}_{c} = ((W_{1,e} + W_{2,e})Acc_{e,1} + W_{3,e}Acc_{e,2}) \\\times P(b \neq 1)
\end{aligned}
\end{equation}
\noindent where $Acc_{e,1}$ denotes the classification accuracy of ${a}_{e,fc} = k$ or ${a}_{e,bp} = k$; $k =$ is the true label.
\begin{equation}
{Acc}_{e,1} = P((a_{e,fc} = k) \cup (a_{e,bp} = k))
\end{equation}
\noindent where $Acc_{e,2}$ denotes the classification accuracy of actions decoded from the fusion of the classification posterior probability ${a}_{e,pp} = k$.
\begin{equation}
{Acc}_{e,2} = P(a_{e,pp} = k)
\end{equation}

\section{Results}\label{sec3}

To evaluate the performance of the implemented copilot models, two simulated environments were explored by the player agents using six different control schemes.
\begin{enumerate}
\item Twin delayed DDPG (TD3): the player agent is controlled solely by the actions of the TD3 actor ($a_{r}$); 
\item Individual EEG (EEG-NB): the player agent is controlled solely by the actions decoded from the fusion of the classification posterior probability of FC and band power (BP) features ($a_{e,pp}$); 
\item EEG-FB: the player agent is controlled by the actions decoded from the fusion of the classification posterior probability of FC and BP features ($a_{e,pp}$), and all risky actions are halted by a blocker ($\beta$); 
\item Co-NB: the player agent is controlled by the actions of the TD3 actor ($a_{r}$) and the actions decoded from human EEG signals ($a_{e}$), without intervention from the blocker ($\beta$); 
\item Co-PPB: the player agent is controlled by the actions of the TD3 actor ($a_{r}$) and the actions decoded from human EEG signals ($a_{e}$), and only risky actions decoded from the fusion of the classification posterior probability ($a_{e,pp}$) are halted by a blocker ($\beta$); 
\item Co-FB: the player agent is controlled by the actions of the TD3 actor ($a_{r}$) and the actions decoded from human EEG signals ($a_{e}$), and all risky actions are halted by a blocker ($\beta$).
\end{enumerate}

\subsection{General Evaluation of Model Performance}\label{sec3.1}

\begin{figure}[h]%
\centering
\captionsetup{width=0.9\textwidth}
\includegraphics[width=0.9\textwidth,keepaspectratio]{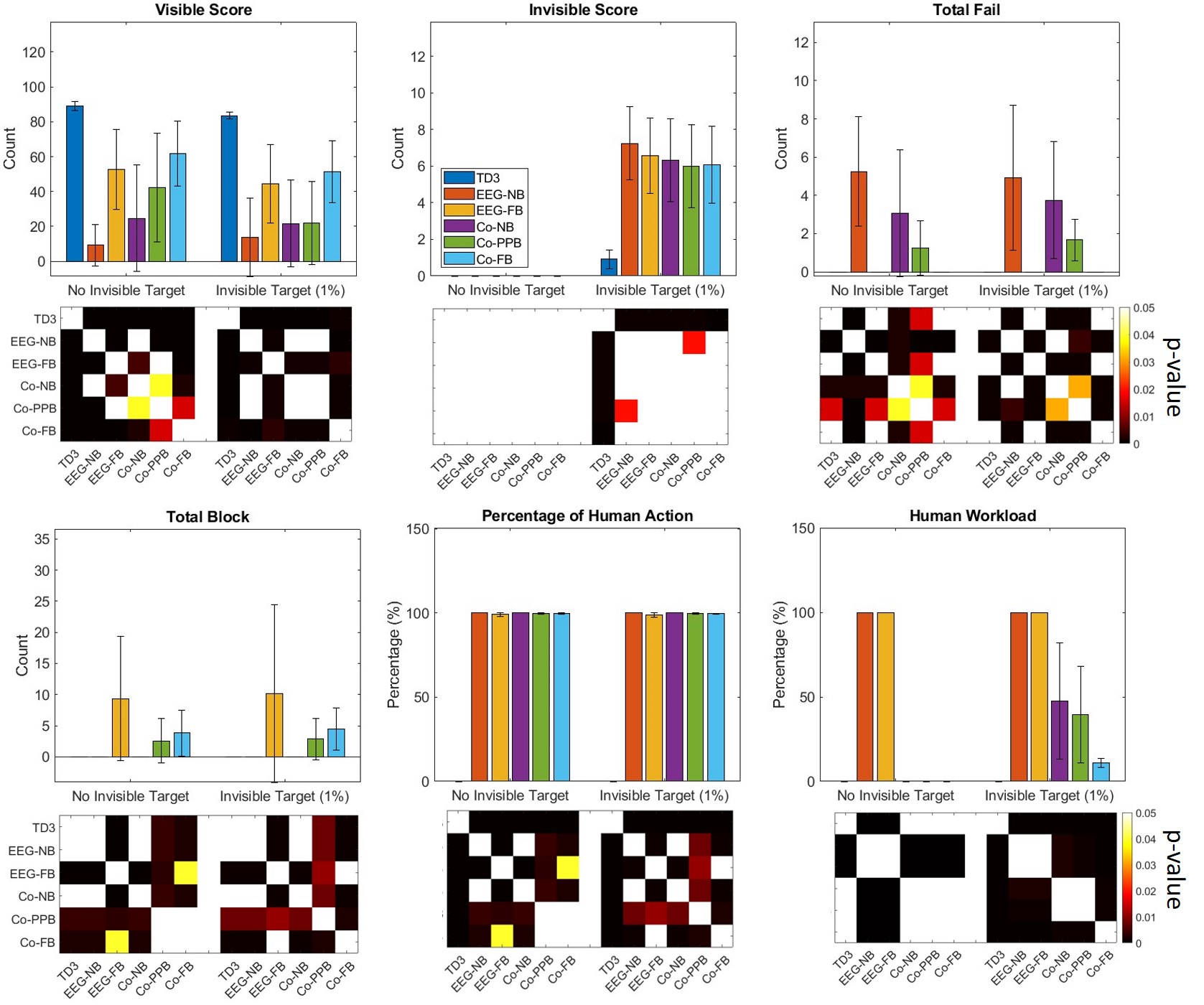}
\caption{Player agents with six different control models explored simulated environments with or without an invisible target. The mean performance metrics of the different models are shown in a bar graph. The respective $p-$values, measured by the Wilcoxon signed-rank test, are shown as a matrix. The copilot models (EEG-FB, Co-NB, Co-PPB, and Co-FB) exhibited promising visible and invisible scores and reduced failure and blockage rates compared with the individual models. In the copilot models, the percentage of human-controllable action remains high, while the human workload remains low.}\label{evaluations}
\end{figure}

Player agents with different control schemes were allowed to explore 1,000 steps in each environment with or without an invisible target. All control schemes were evaluated on the basis of six performance metrics: visible score, invisible score, total fail, total block, percentage of human action, and human workload. For clarity, the visible and invisible scores were measured as the number of visible and invisible targets successfully approached by the player agent. Conversely, total fail and total block were represented by the number of steps that led to game failure and the number of blockages performed by the blocker to prevent risky actions, respectively. The percentage of human action was computed as $100\times\frac{N_{a_e}}{N_{a_e}+N_{a_r}}$, where $N_{a_e}$ denotes the number of actions performed by the human agent and $N_{a_r}$ denotes the number of actions performed by the TD3 agent. Human workload was computed as the number of instances unable to be solved by the TD3 agent, and input from the human agent was required. Human workload was calculated as $100\times\frac{N_{v}}{N_{v}+N_{iv}}$, where $N_{v}$ denotes the number of visible targets approached and $N_{iv}$ denotes the number of invisible targets approached. The percentage of human action measures the maximum allowable intervention, and the human workload measures the minimum state that explicitly requires human intervention.

Figure \ref{evaluations} shows the mean performance across the 12 subjects in the dataset presented by \cite{kaya2018}. The performance was compared across six different control schemes using the Wilcoxon signed-rank test \cite{conover1999}. In the environment with or without an invisible target, the player controlled with the TD3 agent had the best visible score ($1068\pm2.52$ \& $1002\pm1.93$), total fail ($0$), total block ($0$), and human workload ($0\%$). However, this control scheme did not allow input from the human agent ($0\%$) and scored poorly in the invisible score ($11\pm0.52$) in the environment with invisible targets. The invisible score obtained was due to the random appearance of invisible targets. Conversely, the player controlled solely by human EEG (EEG-NB) scored well in the environment with invisible targets ($7.25\pm2.01$); however, its failure rate ($4.92\pm3.80$) and human workload ($100\%$) were suboptimal. The introduction of a blocker network into human EEG control (EEG-FB) significantly improves its visible scores ($52.5\pm22.97$ \& $44.42\pm22.51$) in both environments and reduces the fail count ($0$). However, there was no significant improvement in human workload.

Copilot schemes were proposed to reduce human workload while maintaining the other performance measures. The visible score, invisible score, total fail, and percentage of human action of the EEG-TD3 copilot control scheme without a blocker (Co-NB)  remained similar to those of the EEG-NB model, while the human workload was significantly reduced to $47.71\%\pm34.41\%$. The other posterior probability blocked model (Co-PPB) exhibited very similar performance to the Co-NB model; however, the total fail was significantly decreased to $0$ by the incorporation of the blocker network. The copilot model with full blockage (Co-FB) exhibited promising performance in terms of the six metrics. The visible score of the Co-FB model was significantly improved ($61.75\pm18.66$ \& $51.25\pm17.75$) compared with that of the EEG-based or copilot control schemes, and its invisible score of $6.08\%\pm2.11\%$ remains significantly higher than that of the TD3 control scheme. Player control using the Co-FB model scheme exhibited significantly reduced total fail ($0$) and human workload ($10.81\%\pm2.69\%$) compared with the other models. The percentage of human action ($99.56\%\pm0.34\%$) was also maintained, and the total block remained low at $4.42\%\pm3.40\%$. This indicates that the Co-FB model is the optimal control scheme in our experiment. 

\subsection{Weighted Scores based on humans Interventions}\label{sec3.2}

To better summarize and visualize the performance of the six control schemes, aggregated scores were calculated, as shown in Figure \ref{scores}. The visible score, invisible score, total fail, and total block were weighted and summed as follows:
\begin{equation}
\begin{aligned}
Score_{aggregated} = w_{1}score_{visible} + w_{2}score_{invisible} \\+ w_{3}score_{fail} + w_{4}score_{block}
\end{aligned}
\end{equation}

\noindent where $w$ represents the weights of scores and the reward $r$ provided by the environment. The $w_{1}=1$ is the reward for approaching a visible target, and $w_{2}=10$ is the reward for reaching an invisible target. However, $w_{3}=-10$ is the penalty for failing in the environment, and $w_{4}=-5$ is the penalty for risky but blocked behavior. The percentage of human action is important to allow human authority within the control scheme, such as to deal with emergencies. Human workload is also a crucial measure for disburdening the human agent within the control scheme. Therefore, the aggregated score was further weighted by the percentage of human action and human workload as follows:

\begin{figure}[!h]%
\centering
\captionsetup{width=0.9\textwidth}
\includegraphics[width=0.9\textwidth]{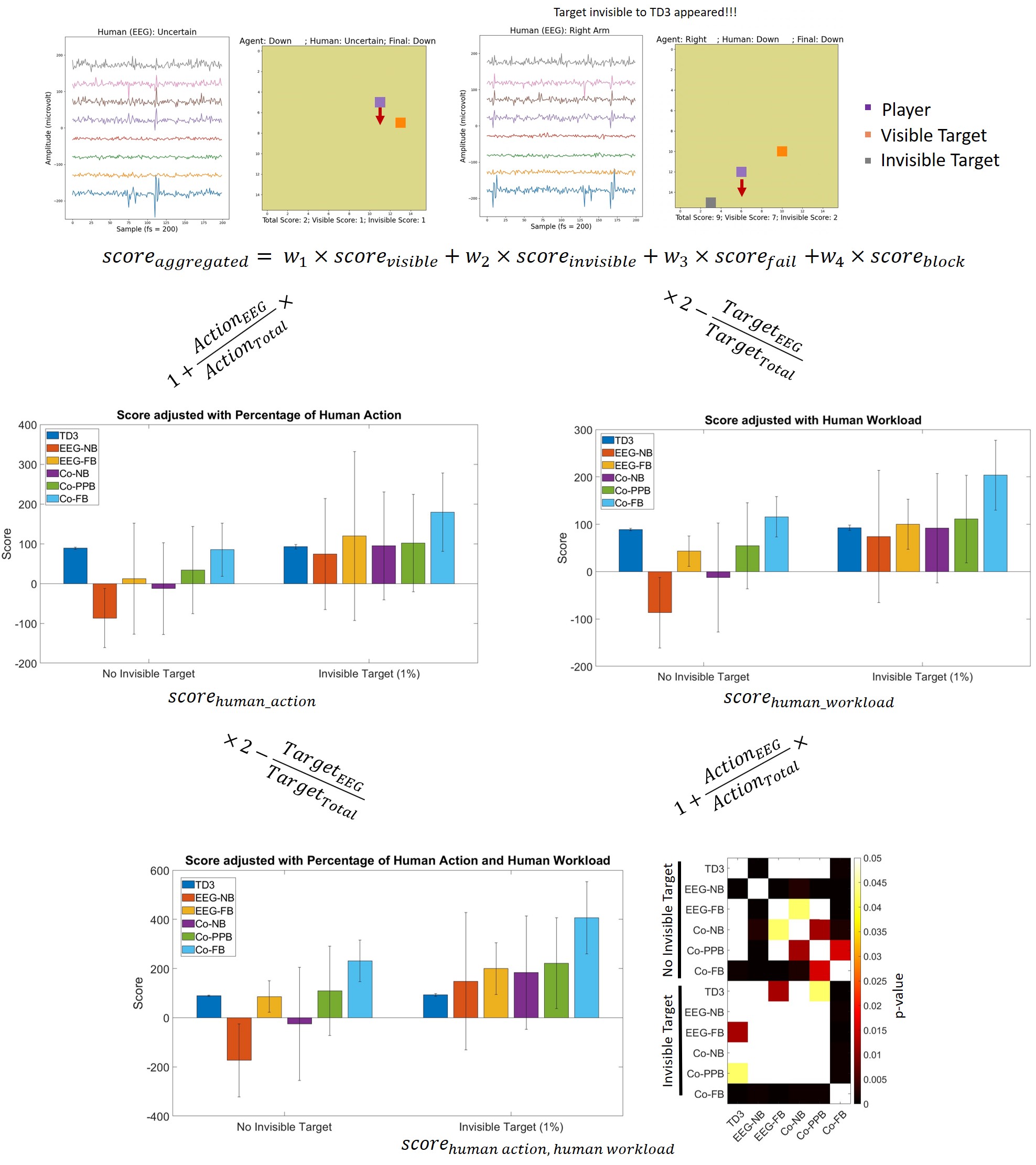}
\caption{Raw visible and invisible scores were weighted positively, and failure and blockage were weighted negatively, according to the rewards from the simulated environment. The aggregated scores were then weighted by the percentage of human action and human workload. Statistical tests of difference are computed using the Wilcoxon signed-rank test and are shown as a matrix. The Co-FB copilot control scheme exhibited significantly higher performance than the remaining five control models.}\label{scores}
\end{figure}

\begin{equation}\label{eq:SHA}
\begin{aligned}
Score_{human\verb!_!action} = score_{aggregated} \\\times (1 + \%_{Human\verb!_!Action})
\end{aligned}
\end{equation}
\begin{equation}\label{eq:SHW}
\begin{aligned}
Score_{human\verb!_!workload} = score_{aggregated} \\\times (2 - \%_{Human\verb!_!Workload})
\end{aligned}
\end{equation}
\noindent The addition of $1$ in Equation \ref{eq:SHA} and the subtraction of $2$ in Equation \ref{eq:SHW} were aimed at restricting $0$ value in the calculation and preventing the score of $0$. Subsequently, a final score was calculated by merging $score_{human\verb!_!action}$ and $score_{human\verb!_!workload}$.

Figure \ref{scores} shows that the scores adjusted by the percentage of human action and human workload of the Co-FB model were higher than those of the other control schemes in the environment with invisible targets . By considering both the percentage of human action and human workload, the Co-FB control scheme significantly outperformed (Wilcoxon signed-rank test, $p<0.001$) the other five models in both environments with or without invisible targets, reporting an adjusted score of $406.66\pm147.28$. Conversely, the adjusted scores of EEG-NB $148.33\pm279.19$ were lower because of the thigh human workload and total failure. Despite its high score for visible targets, the TD3 control scheme was also suboptimal because of its low invisible score and percentage of human action, with a lower adjusted score of $92.67\pm5.42$. We also observed that the incorporation of a full blocker (FB) could drastically improve the performance of both EEG-based (EEG-NB) and copilot (Co-NB, Co-PPB) models, indicating the importance of blocker networks to hinder performance decrease caused by erroneous actions. 

\subsection{Classification Accuracy and Control Authority of Co-FB Model}\label{sec3.3}

The Co-FB model was found to be the best control scheme, as discussed in section \ref{sec3.2}. We further evaluated how the control accuracy of the EEG feature classification was affected by the Co-FB control scheme. The control authorities of the human and TD3 agents were also investigated. Detailed computations of control accuracy and authority are presented in section \ref{sec2.6}.

\begin{figure}[h]%
\centering
\captionsetup{width=0.9\textwidth}
\includegraphics[width=0.7\textwidth]{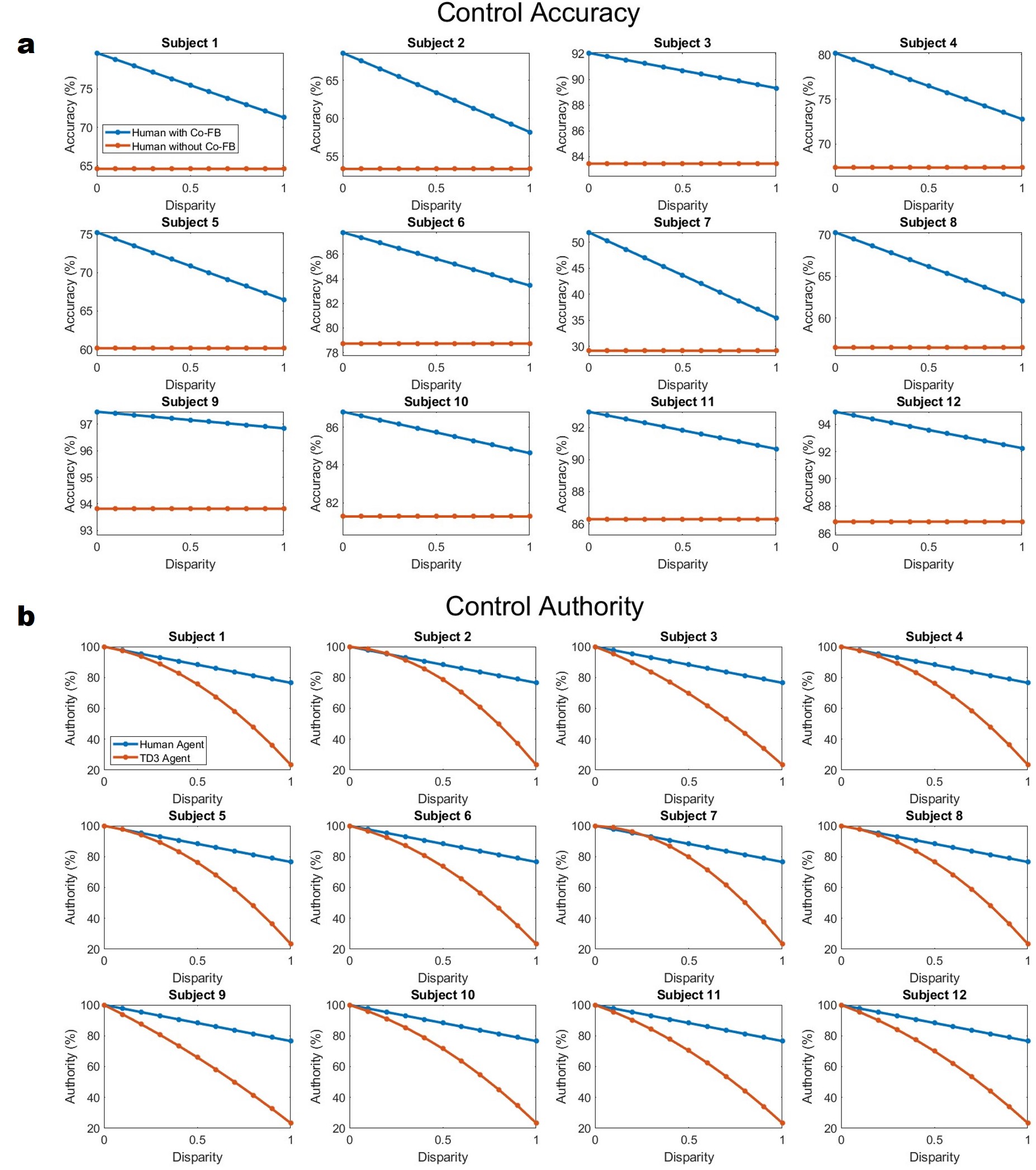}
\caption{With respect to the disparity index, (a) the control accuracy of the human agent with or without the Co-FB scheme, and (b) the control authority of the human agent and TD3 agent in the Co-FB scheme. The control accuracies of all subjects were higher with the Co-FB scheme, and the accuracies decreased as the disparity increased. The control authority of both human and TD3 agents decreases as the disparity increases, with the human agent retaining more authority than the TD3 agent.}\label{accuracy_authority}
\end{figure}

Figure \ref{accuracy_authority} (a) visualizes the control accuracy of a human agent with or without the Co-FB scheme, which is affected by the change in the disparity level. The control accuracy without the Co-FB scheme was the accuracy of merged posterior probabilities classified from EEG FC and BP using a linear discriminant analysis (LDA) classifier. The disparity level did not affect accuracy because no TD3 agent was involved. The control accuracy with the Co-FB scheme demonstrates the classification accuracy of EEG features from a human agent with the assistance of a TD3 agent based on the current state. In this case, accuracy could be affected by the disparity level. We observed that a lower disparity level yielded better classification performance, and the accuracy dropped slightly with an increase in the disparity level. The average improvement dropped from $11.37\%\pm5.61\%$ with $d=0$ to $5.16\%\pm1.67\%$ with $d=1$. However, the accuracy with the Co-FB scheme was higher than that without Co-FB in all 12 subjects, suggesting an improvement in the decoding of EEG intentions with the TD3 agent. We also observed that subjects with lower posterior probability classification accuracy ($Acc_{pp}$) tended to demonstrate higher accuracy improvement $\Delta Acc$ with the incorporation of the copilot scheme. Taking a disparity of 0.5 as an example, the $Acc_{pp}$ of subject 7 was low at $29.14\%$; however, with the assistance of the TD3 agent, the accuracy improved by $14.54\%$. Conversely, subject 9 with $Acc_{pp}$ of $93.82\%$ only exhibited an improvement of $3.33\%$ with the TD3 agent. We found that $Acc_{pp}$ and $\Delta Acc$ were significantly correlated ($p<0.05$) within the disparity range of $d=[0, 1]$, with a median correlation of $-0.94$ and $d=0.5$. This indicated that the collaboration between the human and TD3 agents contributed positively to the classification of EEG features.
	
The control authority of the human and TD3 agents with respect to different disparity indexes is shown in Figure \ref{accuracy_authority} (b). In the case of low disparity, both agents shared high authority in performing the target task. The authority of both agents decreased as the disparity index increased, and the authority of the TD3 agent presented a steeper reduction than that of the human agent. The average difference in control authority between the human and TD3 agents was low at $0.93\%$ when $d=0$, and the mean difference increased to $49.33\%$ when $d$ was increased to $1$. An interesting phenomenon was observed when the authority of the TD3 agent tended to persist higher in subjects with lower EEG classification accuracy ($Acc_{pp}$), despite the increase in disparity. As examples, the $Acc_{pp}$ values of subjects 2 and 7 were $53.35\%$ and $29.14\%$, respectively. The TD3 agents of these subjects exhibited slightly higher control authority than their respective human agents when the disparity level was within the range of $d=[0.1, 0.2]$. This suggests that the TD3 agent can exert higher control authority in a situation where the disparity and EEG classification accuracy are low to improve the overall performance of the Co-FB model. We also found a significant correlation of $-0.93$ ($p<0.001$) between $Acc_{pp}$ and the control authority of the TD3 agent, whereas no significant correlation was found with the control authority of the human agent. 

\subsection{Model Generalizability on BCI Competition Dataset}\label{sec3.4}

\begin{figure*}[h]%
\centering
\captionsetup{width=0.9\textwidth}
\includegraphics[width=0.65\textwidth]{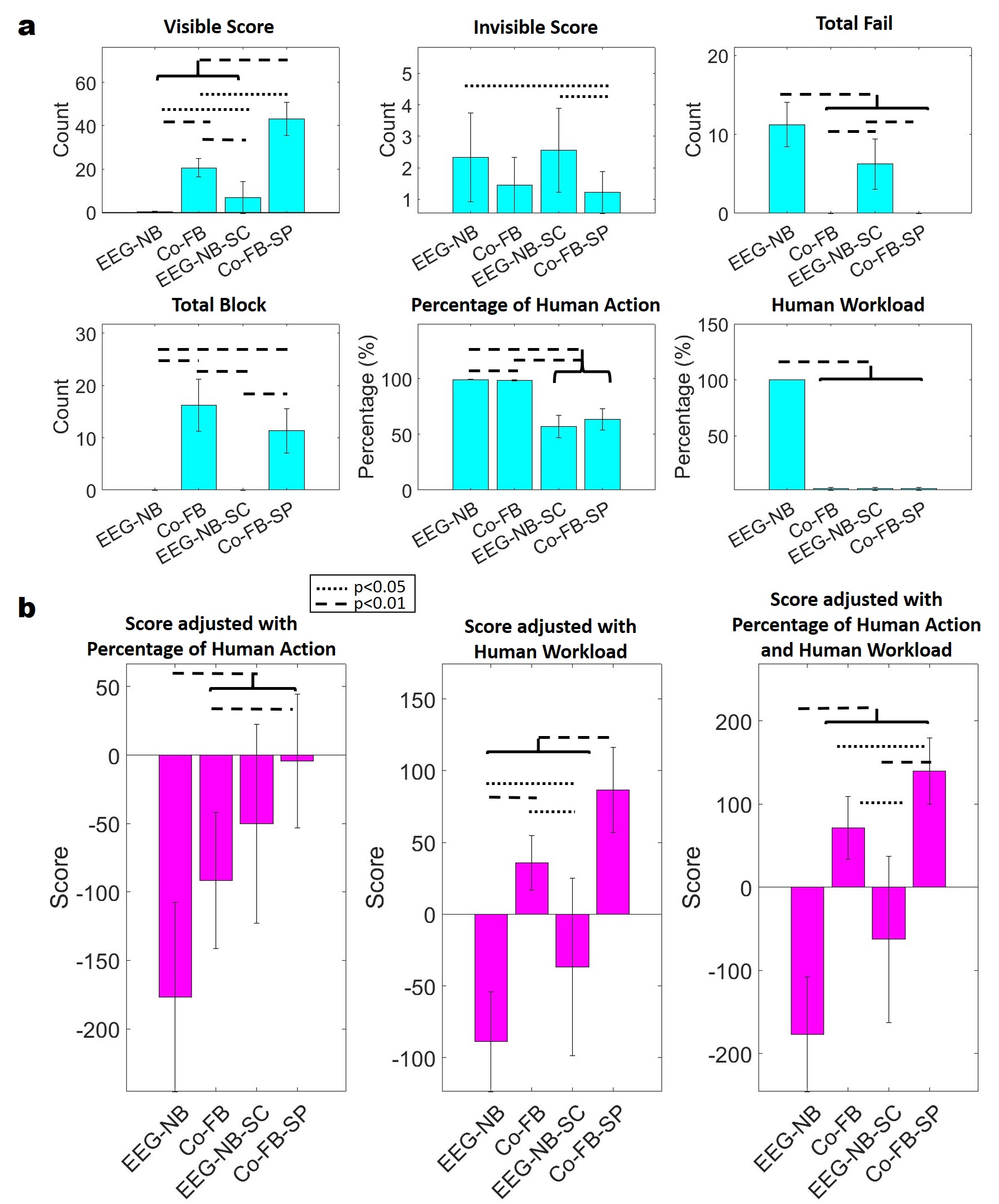}
\caption{Performance of the proposed copilot control scheme with additional BCI Competition EEG dataset, where (a) shows individual scores and (b) shows weighted scores. Because of the lower distinguishability of MI activities in this dataset, the control authority was shifted toward the TD3 agent to improve model performance. The modified model Co-FB-SP exhibits improved performance compared with the other control models, as demonstrated by the score adjusted for the percentage of human action and human workload.}\label{new_dataset}
\end{figure*}

The performance of our proposed copilot approach was further verified with an additional EEG dataset previously published in BCI Competition \cite{tangermann2012}. This EEG dataset demonstrated lower control accuracy without the copilot scheme and reported an average four-class classification accuracy of 31.55\%. Figure \ref{new_dataset} shows the performance of the different copilot schemes on this EEG dataset. Similar to the results found in the previous dataset, we observed that the Co-FB control scheme scored significantly higher in visible score ($20.44\pm4.22$) and lower in total fail ($0$) and human workload ($2.73\%\pm1.13\%$) than the EEG-NB model. However, because of the low EEG classification accuracy, the visible and invisible scores were lower than those in the previous dataset, and the total block was higher. Consequently, although the adjusted scores of the Co-FB model ($71.18\pm37.76$) in this EEG dataset were significantly higher than those of the EEG-NB model ($-176.77\pm69.02$), they did not match the performance of the previous EEG dataset. 

To improve the copilot performance with poorer EEG detection accuracy, we shifted the control authority toward the TD3 agent through a slight modification of the copilot control scheme. The modified Co-FB model is denoted as Co-FB-SP, where the human agent remains semi-passive and the TD3 agent has full control authority when the invisible target is absent. Upon receipt of human input, in situations such as the presence of an invisible target, the Co-FB-SP is introduced into the Co-FB scheme, assisting the human agent in approaching targets while minimizing the risk caused by the misclassification of human EEG features. The Co-FB-SP scheme is reactivated in the absence of human input, returning the control authority to the TD3 agent. For comparison, the performance of the conventional switched control scheme (EEG-NB-SC) was also demonstrated. The EEG-NB-SC scheme, which is similar to the Co-FB-SP scheme, allows full TD3 authority in the absence of human input but shifts to full human authority once the input from humans is detected. 

As shown in Figure \ref{new_dataset} (a), the modified control scheme with semi-passive human input (Co-FB-SP) significantly outperformed Co-FB in terms of visible score ($43.11\pm7.74$), while maintaining the invisible score ($1.22\pm0.67$), total fail ($0$), and total block ($11.33\pm4.24$). However, the percentage of human action was significantly reduced to ($63.43\%\pm9.46\%$), as more authority was given to the TD3 agent in the absence of an invisible target. Compared with the EEG-NB-SC control scheme, Co-FB-SP exhibited better performance in terms of visible score and total fail, and similar performance in terms of the percentage of human action and human workload. The Co-FB-SP model achieved higher adjusted scores ($139.53\pm39.71$) than the other three control schemes (EEG-NB, Co-FB, and EEG-NB-SC), suggesting that shifting control authority is a feasible way to improve the copilot model when the EEG classification accuracy is suboptimal. 

\section{Discussion and Conclusion}\label{sec4}

In this study, we presented shared autonomy schemes between EEG-based BCI and deep RL using a multi-agent copilot approach. In addition, we proposed a disparity $d$-index to measure the rate of disagreement between the human and TD3 agents and to study its effect on model performance. We demonstrated that the Co-FB control scheme allowed active human intervention during the task execution phase of deep RL. According to our hypothesis, the implemented Co-FB model exhibited promising improvements in performance during the execution of complex tasks. Compared with the EEG-NB control scheme, the Co-FB model exhibited improvement in the target approaching score and failure rate while allowing a high level of human intervention and reduced human workload. The Co-FB control scheme outperformed the TD3 model in terms of invisible target approaching and level of human intervention. We also found that the TD3 agent tended to improve the classification performance of the MI-based BCI. The enhancement was likely due to the incorporation of environmental information as an auxiliary parameter to predict MI responses by the TD3 agent trained to navigate the environment. The Co-FB control scheme exhibited generalizability to an additional MI-BCI dataset with lower LDA classification accuracy. However, in the case with low BCI accuracy, we found that shifting more control authority to the TD3 agent increased the adjusted score. 

Because of the difficulty in decoding continuous intention from EEG signals, most BCI studies employ discrete classification. This study is limited to discrete action space, and one of the main strengths of TD3 in continuous action space is yet to be demonstrated using the proposed copilot schemes. In this study, we adopted computationally efficient LDA as a classifier for MI-BCI. This limitation can be overcome using robust classifiers such as FBCSP \cite{ang2008} and other deep learning approaches. Future studies can also evaluate copilot control schemes with various types of BCI (including P300, SSVEP, or speech imagery) and other interesting environments (i.e., robotic control or board game). The performance of copilot control can be enhanced in future studies using different modeling approaches for multi-agent interactions.

\section{Acknowledgments}
This study was partly supported by KAKENHI 21H04956.

\bibliographystyle{IEEEbib_2names}
\bibliography{ref}

\begin{thebibliography}{10}

\bibitem{Li2010}
Y.~Li, et~al.,
\newblock ``An {EEG}-based {BCI} system for {2-D} cursor control by combining {Mu/Beta} rhythm and {P300} potential,''
\newblock {\em {IEEE Transactions on Biomedical Engineering}}, vol. 57, no. 10, pp. 2495--2505, 2010.

\bibitem{Long2011}
J.~Long, et~al.,
\newblock ``Target selection with hybrid feature for {BCI}-based {2-D} cursor control,''
\newblock {\em {IEEE Transactions on Biomedical Engineering}}, vol. 59, no. 1, pp. 132--140, 2011.

\bibitem{Allison2012}
B.~Z. Allison, et~al.,
\newblock ``A hybrid {ERD/SSVEP} {BCI} for continuous simultaneous two dimensional cursor control,''
\newblock {\em {Journal of Neuroscience Methods}}, vol. 209, no. 2, pp. 299--307, 2012.

\bibitem{Li2013}
Y.~Li, et~al.,
\newblock ``A hybrid {BCI} system combining {P300} and {SSVEP} and its application to wheelchair control,''
\newblock {\em {IEEE Transactions on Biomedical Engineering}}, vol. 60, no. 11, pp. 3156--3166, 2013.

\bibitem{Carlson2013}
T.~Carlson and J.~d.~R. Millan,
\newblock ``Brain-controlled wheelchairs: a robotic architecture,''
\newblock {\em IEEE Robotics \& Automation Magazine}, vol. 20, no. 1, pp. 65--73, 2013.

\bibitem{Frisoli2012}
A.~Frisoli, et~al.,
\newblock ``A new gaze-{BCI}-driven control of an upper limb exoskeleton for rehabilitation in real-world tasks,''
\newblock {\em {IEEE Transactions on Systems, Man, and Cybernetics, Part C (Applications and Reviews)}}, vol. 42, no. 6, pp. 1169--1179, 2012.

\bibitem{Barsotti2015}
M.~Barsotti, et~al.,
\newblock ``A full upper limb robotic exoskeleton for reaching and grasping rehabilitation triggered by {MI-BCI},''
\newblock {\em {2015 IEEE International Conference on Rehabilitation Robotics (ICORR)}}, pp. 49--54, 2015.

\bibitem{Frolov2016}
A.~Frolov, et~al.,
\newblock ``Preliminary results of a controlled study of {BCI}-exoskeleton technology efficacy in patients with poststroke arm paresis,''
\newblock {\em {Bulletin of Russian State Medical University}}, , no. 2, 2016.

\bibitem{Decety1990}
J.~Decety and D.~H. Ingvar,
\newblock ``Brain structures participating in mental simulation of motor behavior: {A} neuropsychological interpretation,''
\newblock {\em {Acta Psychologica}}, vol. 73, no. 1, pp. 13--34, 1990.

\bibitem{Nielsen2006}
K.~D. Nielsen, et~al.,
\newblock ``Eeg based bci-towards a better control. brain-computer interface research at aalborg university,''
\newblock {\em {IEEE Transactions on Neural Systems and Rehabilitation Engineering}}, vol. 14, no. 2, pp. 202--204, 2006.

\bibitem{Liu2019}
Y.-H. Liu, et~al.,
\newblock ``Analysis of electroencephalography event-related desynchronisation and synchronisation induced by lower-limb stepping motor imagery,''
\newblock {\em {Journal of Medical and Biological Engineering}}, vol. 39, no. 1, pp. 54--69, 2019.

\bibitem{Sun2013}
M.~Sun, et~al.,
\newblock ``{Asynchronous brain-computer interface with foot motor imagery},''
\newblock {\em {2013 ICME International Conference on Complex Medical Engineering, CME 2013}}, pp. 191--196, 2013.

\bibitem{Hashimoto2013}
Y.~Hashimoto and J.~Ushiba,
\newblock ``{EEG-based classification of imaginary left and right foot movements using beta rebound},''
\newblock {\em {Clinical Neurophysiology}}, vol. 124, no. 11, pp. 2153--2160, 2013.

\bibitem{Mohamed2018}
E.~A. Mohamed, et~al.,
\newblock ``Comparison of {EEG} signal decomposition methods in classification of motor-imagery {BCI},''
\newblock {\em {Multimedia Tools and Applications}}, vol. 77, no. 16, pp. 21305--21327, 2018.

\bibitem{ang2008}
K.~K. Ang, et~al.,
\newblock ``{Filter bank common spatial pattern (FBCSP) in brain-computer interface},''
\newblock {\em 2008 IEEE international joint conference on neural networks (IEEE world congress on computational intelligence)}, pp. 2390--2397, 2008.

\bibitem{phang2022}
C.-R. Phang, et~al.,
\newblock ``Frontoparietal dysconnection in covert bipedal activity for enhancing the performance of the motor preparation-based brain--computer interface,''
\newblock {\em IEEE Transactions on Neural Systems and Rehabilitation Engineering}, vol. 31, pp. 139--149, 2022.

\bibitem{Blankertz2010}
B.~Blankertz, et~al.,
\newblock ``Neurophysiological predictor of {SMR}-based {BCI} performance,''
\newblock {\em Neuroimage}, vol. 51, no. 4, pp. 1303--1309, 2010.

\bibitem{Vidaurre2010}
C.~Vidaurre and B.~Blankertz,
\newblock ``Towards a cure for {BCI} illiteracy,''
\newblock {\em Brain topography}, vol. 23, no. 2, pp. 194--198, 2010.

\bibitem{bharadhwaj2020}
H.~Bharadhwaj, et~al.,
\newblock ``Conservative safety critics for exploration,''
\newblock {\em arXiv preprint arXiv:2010.14497}, 2020.

\bibitem{li2022}
Q.~Li, et~al.,
\newblock ``Efficient learning of safe driving policy via human-ai copilot optimization,''
\newblock {\em arXiv preprint arXiv:2202.10341}, 2022.

\bibitem{xu2021}
D.~Xu, et~al.,
\newblock ``Accelerating reinforcement learning using eeg-based implicit human feedback,''
\newblock {\em Neurocomputing}, vol. 460, pp. 139--153, 2021.

\bibitem{rea2022}
D.~J. Rea and S.~H. Seo,
\newblock ``Still not solved: A call for renewed focus on user-centered teleoperation interfaces,''
\newblock {\em Frontiers in Robotics and AI}, vol. 9, pp. 704225, 2022.

\bibitem{yang2021}
C.~Yang, et~al.,
\newblock ``A review of human--machine cooperation in the robotics domain,''
\newblock {\em IEEE Transactions on Human-Machine Systems}, vol. 52, no. 1, pp. 12--25, 2021.

\bibitem{flemisch2019}
F.~Flemisch, et~al.,
\newblock ``Joining the blunt and the pointy end of the spear: towards a common framework of joint action, human--machine cooperation, cooperative guidance and control, shared, traded and supervisory control,''
\newblock {\em Cognition, Technology \& Work}, vol. 21, pp. 555--568, 2019.

\bibitem{schwarting2017}
W.~Schwarting, et~al.,
\newblock ``Parallel autonomy in automated vehicles: Safe motion generation with minimal intervention,''
\newblock {\em 2017 IEEE International Conference on Robotics and Automation (ICRA)}, pp. 1928--1935, 2017.

\bibitem{mars2014}
F.~Mars, et~al.,
\newblock ``Analysis of human-machine cooperation when driving with different degrees of haptic shared control,''
\newblock {\em IEEE transactions on haptics}, vol. 7, no. 3, pp. 324--333, 2014.

\bibitem{reddy2018}
S.~Reddy, et~al.,
\newblock ``Shared autonomy via deep reinforcement learning,''
\newblock {\em arXiv preprint arXiv:1802.01744}, 2018.

\bibitem{wang2022}
X.~Wang, et~al.,
\newblock ``Error-related potential-based shared autonomy via deep recurrent reinforcement learning,''
\newblock {\em Journal of Neural Engineering}, vol. 19, no. 6, pp. 066023, 2022.

\bibitem{kaya2018}
M.~Kaya, et~al.,
\newblock ``A large electroencephalographic motor imagery dataset for electroencephalographic brain computer interfaces,''
\newblock {\em Scientific data}, vol. 5, no. 1, pp. 1--16, 2018.

\bibitem{tangermann2012}
M.~Tangermann, et~al.,
\newblock ``Review of the bci competition iv,''
\newblock {\em Frontiers in neuroscience}, p.~55, 2012.

\bibitem{fujimoto2018}
S.~Fujimoto, et~al.,
\newblock ``Addressing function approximation error in actor-critic methods,''
\newblock {\em International conference on machine learning}, pp. 1587--1596, 2018.

\bibitem{saunders2017}
W.~Saunders, et~al.,
\newblock ``Trial without error: Towards safe reinforcement learning via human intervention,''
\newblock {\em arXiv preprint arXiv:1707.05173}, 2017.

\bibitem{prakash2019}
B.~Prakash, et~al.,
\newblock ``Improving safety in reinforcement learning using model-based architectures and human intervention,''
\newblock {\em arXiv preprint arXiv:1903.09328}, 2019.

\bibitem{van2016}
H.~Van~Hasselt, et~al.,
\newblock ``Deep reinforcement learning with double q-learning,''
\newblock {\em Proceedings of the AAAI conference on artificial intelligence}, vol. 30, no. 1, 2016.

\bibitem{kingma2014}
D.~P. Kingma and J.~Ba,
\newblock ``Adam: A method for stochastic optimization,''
\newblock {\em arXiv preprint arXiv:1412.6980}, 2014.

\bibitem{conover1999}
W.~J. Conover,
\newblock {\em Practical nonparametric statistics}, vol. 350,
\newblock john wiley \& sons, 1999.

\end{thebibliography}

\end{document}